\documentclass{aa}
\usepackage{graphicx}
\usepackage{txfonts}
\begin{document}

   \title{Speeding up the GENGA N-body integrator on consumer-grade graphics cards}

   \author{R. Brasser\inst{1}\fnmsep\thanks{rbrasser@konkoly.hu},
          S. L. Grimm\inst{2,3},
          P. Hatalova\inst{4}
        \and
          J. G. Stadel\inst{5}
          }
   \institute{Research Centre for Astronomy and Earth Sciences; Konkoly Thege Miklos St. 15-17, H-1121 Budapest, Hungary
        \and
        University of Bern, Physikalisches Institut; Gesellschaftsstrasse 6, CH-3012 Bern, Switzerland
        \and
        ETH Z\"{u}rich, Institute for Particle Physics and Astrophysics; Wolfgang-Pauli-Strasse 27, CH-8093 Z\"{u}rich, Switzerland
        \and
        Centre for Planetary Habitability (PHAB), University of Oslo, N-0315 Oslo, Norway
        \and
        University of Z\"{u}rich, Institute for Computational Science; Winterthurerstrasse 190, CH-8057 Z\"{u}rich, Switzerland}

\authorrunning{Brasser et al.}
   \date{Accepted 5 September 2023}

 
  \abstract
{{ Graphics processing unit (GPU) computing has become popular due to the enormous calculation potential that can be harvested from a single card. The N-body integrator Gravitational ENcounters with GPU Acceleration (GENGA) is built to harvest the computing power from such cards, but it suffers a severe performance penalty on consumer-grade Nvidia GPUs due to their artificially truncated double precision performance.}}
{We aim to speed up GENGA on consumer-grade cards by harvesting their high single-precision performance.}
{We modified GENGA to have the option to compute the mutual long-distance forces between bodies in single precision and tested this with five experiments. First, we ran a high number of simulations with similar initial conditions of on average 6600 fully self-gravitating planetesimals in both single and double precision to establish whether the outcomes were statistically different. These simulations were run on Tesla K20 cards. { We supplemented this test with simulations that i) began with a mixture of planetesimals and planetary embryos, ii) planetesimal-driven giant planet migration, and iii) terrestrial planet formation with a dissipating gas disc. All of these simulations served to determine the accuracy of energy and angular momentum conservation under various scenarios with single and double precision forces.} Second, we ran the same simulation beginning with 40\,000 self-gravitating planetesimals using both single and double precision forces on a variety of consumer-grade and Tesla GPUs to measure the performance boost of computing the long-range forces in single precision.}
{{ We find that there are no statistical differences when simulations are run with the gravitational forces in single or double precision that can be attributed to the force prescription rather than stochastic effects. The accumulations in uncertainty in energy are almost identical when running with single or double precision long-range forces. However, the uncertainty in the angular momentum using single rather than double precision long-range forces is about two orders of magnitude greater, but still very low.} Running the simulations in single precision on consumer-grade cards decreases running time by a factor of three and becomes within a factor of three of a Tesla A100 GPU. Additional tuning speeds up the simulation by a factor of two across all types of cards.}
{The option to compute the long-range forces in single precision in GENGA when using consumer-grade GPUs dramatically improves performance at a little penalty to accuracy. { There is an additional environmental benefit because it reduces energy usage.}}


   \maketitle
%
\section{Introduction}
The Gravitational ENcounters with GPU Acceleration (GENGA) N-body code \citep{grimmstadel2014,grimmetal2022} was developed to tackle the next generation of N-body problems in planetary science by harvesting the vast computational power of NVidia Graphics Processing Units (GPUs) and their Compute Unified Device Architecture (CUDA) programming language. On average, a high-end Tesla GPU, such as the V100 or A100, has a comparable amount of computing power to a 128-core AMD Epyc CPU, but with less power consumption \citep{PortegiesZwart2020}. Since the release of the first version of GENGA \citep{grimmstadel2014}, the performance of available computational resources have improved dramatically, especially that of GPUs. On the other hand, the requirements of state-of-the-art simulations in planet formation have increased in step. Current top of the line hardware allows for 30,000 -- 60,000 fully interactive planetesimals to be simulated for 10 million years (Myr) in about 60 days using double precision (FP64 precision). Examples of such high-end simulations are given in \cite{quarleskaib2019}, \cite{clementetal2020}, and \cite{wooetal2021}. \\

\begin{figure}
\resizebox{\hsize}{!}{\includegraphics{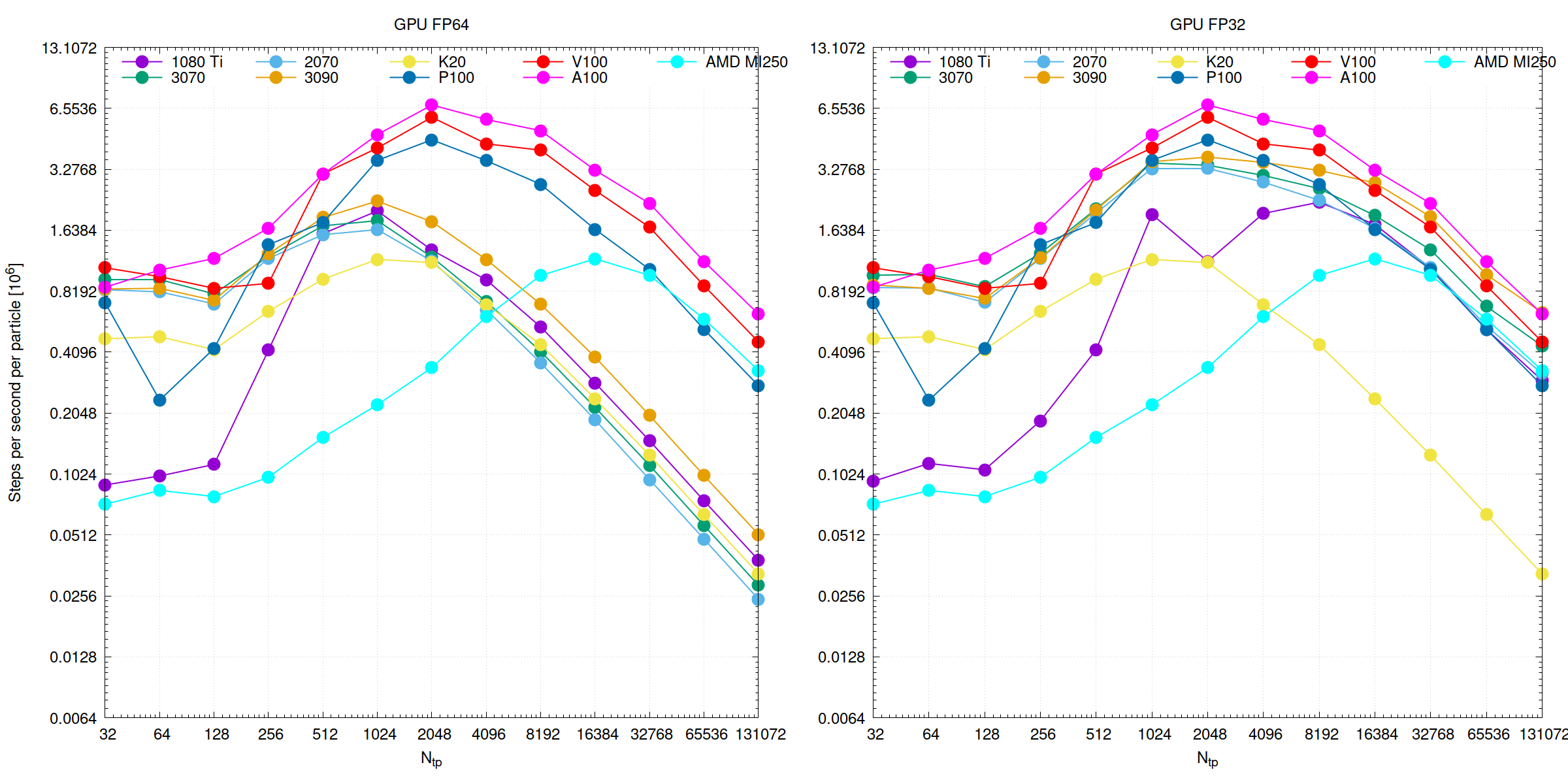}}
\caption{Number of steps per second per planetesimal in GENGA for various GPUs as a function of number of self-gravitating planetesimals. The tests were run with FP64 precision long-range forces with tuning (left panel) and FP32 long-range forces (right panel).}
\label{fig:gpufsg}%
\end{figure}

Yet, such high-N simulations can only be performed within 60 days on the Tesla range of GPUs, which are specifically engineered for computing \citep{Lindholm2008}. Examples of these GPUs are the P100, V100, A100, and the new H100. The performance decreases dramatically when attempting to run the same simulations on consumer-grade GPUs, that is the GTX and RTX series \citep{Mukunoki2016}. The reason for this is clear: the latter cards were created for graphics rendering for computer gaming, and not for computation. In display graphics usage FP64 calculations play only a minor role and the FP64 performance of these cards is truncated to typically 1.6\%--4.2\% of their single-precision (FP32 precision) performance. In order to accurately preserve total energy and total angular momentum in a simulation GENGA predominantly uses FP64 precision, thereby causing a severe performance penalty on consumer-grade GPUs, { and increasing overall power consumption}. For example, for $N\geq 16384$ self-gravitating planetesimals the A100 GPU is about 16 times faster than a GTX 1080 Ti and a RTX 3070 in double precision, with a purchasing price to match. The reason for this discrepancy is mostly due to the truncated FP64 performance of the consumer-grade cards. { In the left panel of Fig.~\ref{fig:gpufsg} we plot the number of steps per second per particle in GENGA for various GPUs as a function of the number of planetesimals using FP64 precision. The enormous difference in speed between the Tesla (P/V/A100) and consumer-grade (GTX/RTX) cards is obvious.} The Tesla K20 has { about half the} clock speed (0.7 GHz) of the Tesla 100 series (up to 1.4 GHz) and the GTX/RTX series (1.5-1.9 GHz), { which explains the lower speed at low $N$}, and a much older architecture { and compute capability, which explains its much slower speed at high $N$.}\\

{ In this work we investigate whether we can substantially speed up GENGA on consumer-grade GPUs by performing the long-range gravitational force calculations, in other words those forces outside of the close encounter phase, in FP32 precision, at the potential cost of higher errors in energy and angular momentum. GENGA's Kepler orbit solver is always performed in double precision, however, to ensure the maximum level of energy conservation. For high $N$ simulations, typically $N\gtrsim 2048$, the calculation of the $N^2$ gravitational force terms are computationally the most expensive part. Therefore, theoretically, a large speed up can be achieved to optimise this part as much as possible. In the right panel of Fig.~\ref{fig:gpufsg} the same tests are performed as on the left panel but with the long-range forces computed in FP32 precision. It is clear that the consumer-grade cards now have comparable performance to the Tesla P100!}\\

GENGA has introduced a hierarchical method that uses multiple changeover functions so that it is able to integrate large close-encounter groups very efficiently \citep{grimmetal2022}. The original hybrid symplectic integrator that GENGA employs consists of two levels. The basic level is the symplectic integrator with a fixed step size wherein bodies revolve around the central mass (level 0), and the Bulirsch-Stoer integrator (level 1), which is used for close-encounters between bodies other than the central mass. A changeover function smoothly switches between the two levels \citep{chambers1999}. This method can be extended by introducing a number of intermediate levels between the fully symplectic regime and the Bulirsch-Stoer regime: instead of switching directly to the Bulirsch-Stoer method, another symplectic integration step with a reduced time step can be applied in between. This is desirable because the Bulirsch-Stoer steps are computationally expensive and should be used as a last resort only when bodies are very close to each other. \\

The limit where the changeover functions are applied is defined by a critical radius $r_{\text{crit}}$ of a particle $i$, which is computed as \citep{grimmetal2022}

\begin{equation}
    \label{rcrit}
    r_{\text{crit},i}= \max(n_1 \cdot R_{H,i}, n_2 \cdot dt \cdot v_i).
\end{equation}

The critical radius depends on two terms. The first depends on the Hill radius { $R_{H,i}=a_i(m_i/3M_*)^{1/3}$, where $m_i$ is the mass of body $i$, $M_*$ the mass of the central body, and $a_i$ is the semi-major axis. The} second term contains the time step $dt$ and the Kepler velocity $v_i$ of the particle $i$. The two parameters $n_1$ and $n_2$ are constants. Experimentation has found that typical values that yield a good compromise between accuracy and computation time are $n_1=3$ and $n_2=0.4$ \citep{chambers1999}.\\

In practice, a good balance between the number of levels and the number of sub-steps must be found. Using a higher number of levels requires more memory to store all close-encounter pairs of each level, and using a higher number of sub-steps increases the amount of CUDA kernel calls needed; this can increase the total kernel overhead time, and thus the total computation time. The configuration of levels and sub-steps that will yield the shortest integration time will depend mostly on the number of bodies and their surface density, and a little on the GPU device used. As such, GENGA now tests different configurations of subsets before running the simulation in order to find the best settings and perform at the fastest speed possible. As the simulation evolves the optimal parameters may change, so that regular halting and resuming of simulations, such as is done on shared High-Performance Computing (HPC) systems, will result in optimal performance.\\

It is important to note that during a close encounter between bodies, the gravitational force between two or more involved particles is always calculated in double precision using changeover functions. Employing the before-mentioned self-tuning step in the beginning of the simulation in combination with frequent stopping and restarting, and employing FP32 precision long-range force calculations yields a total speedup of a factor of six to eight on consumer-based cards than when both of these options are disabled. With single precision long-range forces the speed ratio between consumer-based GPUs and the A100 decreases from $\sim 16$ to about a factor of three.

\section{Methodology}
We performed { two sets of numerical experiments. The first set consists of different types of simulations on the same hardware -- Tesla K20 cards. The aim of these simulations was to establish whether there were large statistical deviations between simulations with the same initial conditions that were run with forces in either FP64 or FP32 precision. These experiments consist of: a) 80 terrestrial planet formation simulations from an annulus of self-gravitating planetesimals without a gas disc, and with Jupiter and Saturn on their current orbits; b) eight simulations of a mixture of planetary embryos and planetesimals with varying mass ratios without a gas disc; c) 56 planetesimal-driven giant planet migration simulations; and d) ten terrestrial planet formation simulations with a gas disc present, and with Jupiter and Saturn on their current orbits. The initial conditions of these last simulations are identical to the low-resolution simulations described in \citet{wooetal2021}. All of these tests present different challenges to the integrator.} \\

The second experiment consists of running the same initial conditions with both FP64 and FP32 precision long-range forces on various GPUs, and establish the rate of computation. This last set was run both with and without the self-tuning step: it was run four times on each type of GPU that we tested. The GPUs that we tested were the Teslas K20, P100, and A100, the GTX 1080 Ti, the RTX 2070, 3070, and 3090, and the AMD Instinct MI250 using the Heterogeneous-Compute Interface for Portability (HIP) version of GENGA \citep{grimmetal2022}. { In all experiments the collisions are assumed to result in perfect mergers.}\\

The first set of simulations has 80 runs with on average 6600 self-gravitating planetesimals situated between 0.7 au and 1 au with a total mass of 2 Earth masses ($M_\oplus$). We included Jupiter and Saturn on their current orbits. Planetesimals had a diameter between 600 km and 2000 km with a cumulative size-frequency distribution slope of $-1.8$ in accordance with planetesimal formation simulations \citep[e.g.][]{Johansen2015}. We ran the simulations with different realisations of the planetesimals on K20 GPUs, with 40 with FP64 long-range force precision, and the other 40 with FP32 long-range force precision. The time step was set to 5 days and the total duration was 10 Myr. Planetesimals were removed once they ventured closer then 0.4 au or farther than 40 au from the Sun, or collided with each other or the giant planets. We did not employ the effects of a gas disc. { The initial conditions of run $i+40$ were identical to run $i$.\\

The second set consisted eight simulations of 10 or 20 Mars-mass planetary embryos and 5000 equal-mass planetesimals with a combined mass of 2~$M_\oplus$. The total mass in embryos and planetesimals was 1:1, 10:1, 30:1, and 100:1. All bodies were situated between 0.7 au and 1.0 au without the giant planets present. This semi-major axis range was chosen to as to facilitate comparison of the outcomes with the first set of simulations. Simulations were run for 10 Myr with a time step of 0.01~yr. Bodies were removed once they ventured closer then 0.3 au or farther than 10 au from the Sun, collided with each other or the giant planets. We did not employ the effects of a gas disc. Four simulations were run in FP64 mode and four in FP32 mode. The initial conditions of run $i+4$ were identical to run $i$.\\

\begin{figure}
\resizebox{\hsize}{!}{\includegraphics{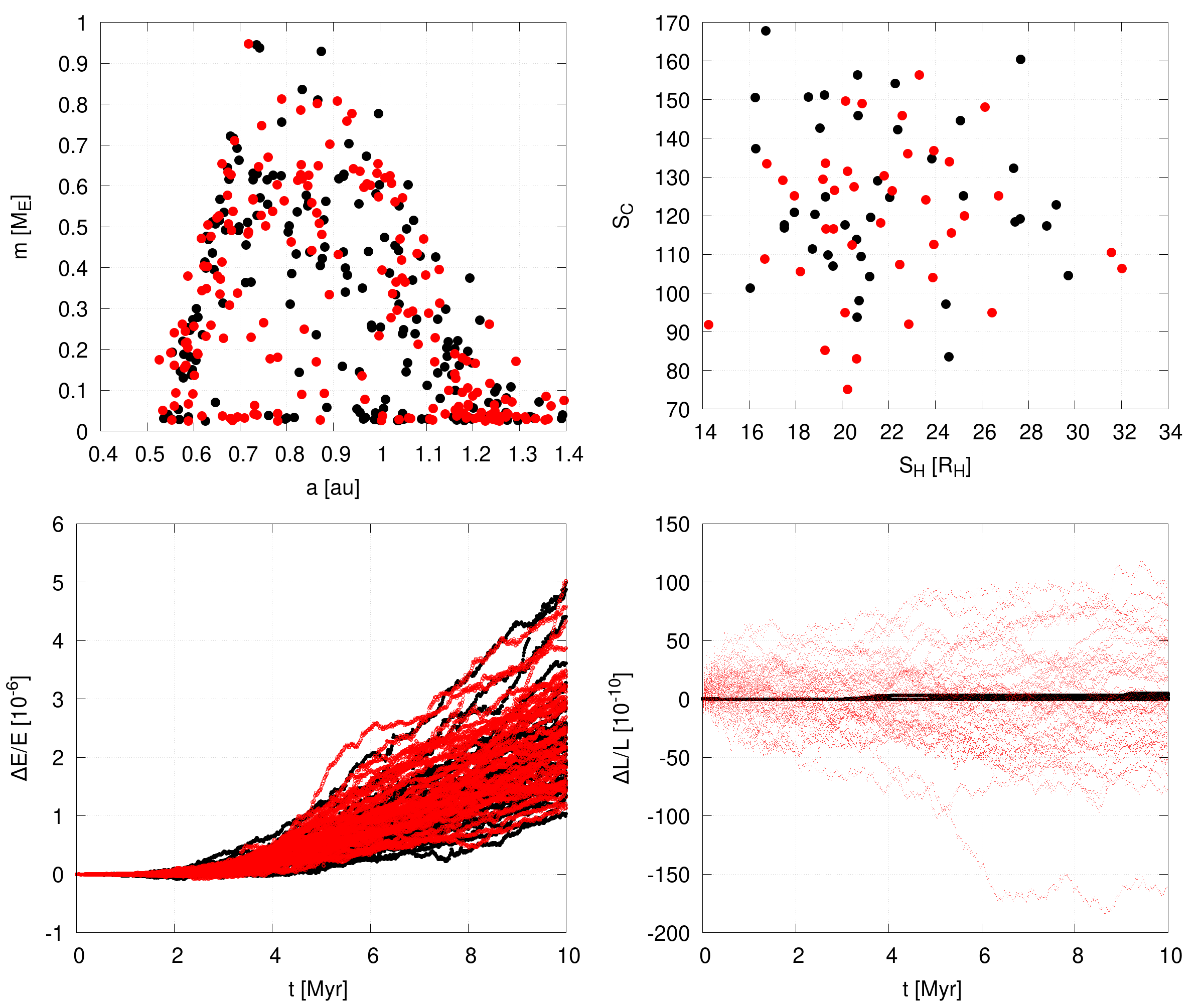}}
\caption{Chambers statistics for the 80 sims from the first experiment. Black dots are for the sims with FP64 precision forces, red dots use FP32 precision forces. The top-left panel shows the final mass versus final semi-major axis for all objects with a final mass $m>0.025$~$M_\oplus$. The top-right panel shows the concentration parameter versus spacing ($S_M$ vs. $S_H$). The bottom-left panel the relative change in energy, and the bottom-right panel depicts the the relative change in angular momentum.}
\label{fig:chambers}%
\end{figure}

\begin{figure*}[ht]
\resizebox{\hsize}{!}{\includegraphics{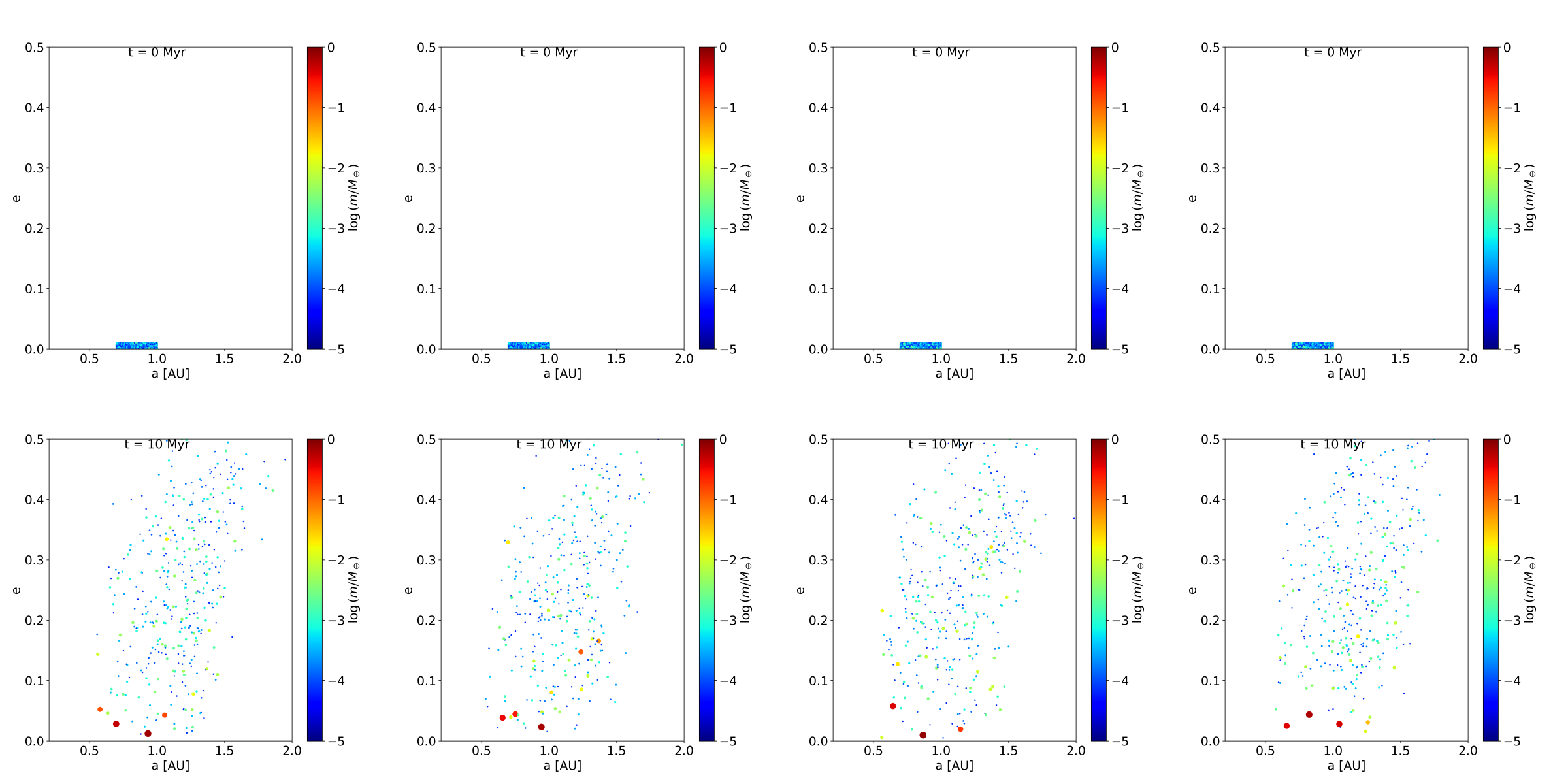}}
\caption{Initial (top row) and final (bottom row) semi-major axes, eccentricities, and masses for four random simulations. The leftmost two are run with long-range forces with FP64 precision while the rightmost two have FP32 precision long-range forces.}
\label{fig:tpfbe}%
\end{figure*}

The third set consisted of 56 simulations of planetesimal-drive giant planet migration. The planets started on the loose five-planet configuration of \citet{NM2012}. The planets were surrounded by a planetesimal disc of 18~$M_\oplus$ consisting of 30\,000 equal-mass bodies with an outer edge of 27~au and an inner edge of 1~au beyond Neptune \citep{Wong2019}. The planetesimal disc was not fully self-gravitating. The simulations were run for 200~Myr with a time step of 0.3~yr. Bodies were removed closer than 1.7~au and farther than 3000~au from the Sun, or when they collided with a planet. Simulations were stopped when fewer than four giant planets remained, or until the end was reached, whichever came first. Twenty-eight simulations were run in FP64 mode and 28 in FP32 mode. The initial conditions of run $i+28$ were identical to run $i$. \\

The fourth set consisted of a rerun of the low-resolution EJS simulations of \citet{wooetal2021} with the long-range forces in FP32 precision. The total number of simulations is ten. The initial conditions, time step, and removal criteria are the same: 5 day time step, 10~Myr simulation duration, minimum-mass solar nebula gas disc (surface density of $\Sigma=1700$~g~cm$^{-2}$ at 1 au, gas surface density slope of $-1$) with a decay e-folding time of 1~Myr, Jupiter and Saturn on their current orbits, and objects were removed when closer than 0.1~au or farther than 100~au from the Sun. {The gas exerts a drag force and type I migration as well as the force of the potential of the gas disc itself on all bodies. Details of the force prescription and implementation can be found in \citet{Morishima2010}, and in \citet{grimmetal2022}.}\\

The last experiment also consisted of 2~$M_\oplus$ of planetesimals situated between 0.7 au and 1 au, but now the minimum and maximum diameter were 200 km and 2000 km, respectively. This resulted in 40322 planetesimals, with the same slope as in the first set. The same initial conditions were run four times on the same GPU: once with FP64 force precision, once with FP32 force precision, and once again with the self-tuning enabled. The initial conditions were the same for all GPUs. These simulations were run for 1 Myr with the same time step and removal criteria as for the first experiment. \\

{ For the first, second, and fourth experiments} we analyse the output after 10 Myr and compute several planetary system quantities introduced by \cite{chambers2001}, which describe the general dynamical properties of a planetary system. The first is the angular momentum deficit (AMD) -- not to be confused with the manufacturer -- which is given by

\begin{equation}
 {\rm AMD} = \frac{\sum_k \mu_k\sqrt{a_k}(1-\sqrt{1-e_k^2})\cos i_k}{\mu_k \sqrt{a_k}}
\end{equation}
where $\mu_k=m_k/M_\odot$ and $m_k$ is the mass of planetesimal $k$, $a_k$ is the semi-major axis, $e_k$ is the eccentricity, and $i_k$ is the inclination. The second is the fraction of mass in the most massive planet ($S_{\rm M}$). The third is a concentration parameter ($S_{\rm C}$), given by

\begin{equation}
S_{\rm C} = {\rm max}\Bigl(\frac{\sum_k \mu_k}{\sum_k \mu_k [\log(a/a_k)]^2}\Bigr),
\end{equation}
and last, a mean spacing parameter ($S_{\rm H}$), which is

\begin{equation}
 S_H = \frac{1}{N-1}\sum_{k=1}^{N-1}\frac{a_{k+1}-a_{k}}{a_{k+1}+a_k}\Bigl(\frac{\mu_{k+1}+\mu_k}{3}\Bigr)^{-1/3}.
\end{equation}
Unlike \cite{chambers2001} we use the mutual Hill sphere as the spacing unit. For all experiments we focus primarily on the evolution of the energy and angular momentum, and whether these stay within reasonable bounds in the FP32 simulations.

\section{Results}
\subsection{Terrestrial planet formation without gas}
{The simulations run for this experiment were performed much earlier than those of the other experiments, and were run with an older version of GENGA. As such, the behaviour in energy and angular momentum is somewhat different from that of the ensuing sets.} \\

In Fig.~\ref{fig:chambers} we plot the mass vs semi-major axis (top-left panel), the { concentration parameter} versus the spacing parameter (top-right), { the temporal evolution of the relative change in energy, $\Delta E/E = E(t)/E_{t=0}-1$} (bottom-left) and { the temporal evolution of the relative change in angular momentum, $\Delta L/L = L(t)/L_{t=0}-1$} (bottom-right). The black dots are for the FP64 simulations, the red dots pertain to the FP32 runs. It is clear that the outcomes of the two sets of simulations are nearly identical. The outcome of these simulations yield planetary systems that are incompatible with the current terrestrial planets due to the short integration time. As the figure shows, running the simulations with the forces in FP32 precision yields statistically identical planetary systems after 10 Myr of evolution. { The relative energy keeps rising, but their values are of identical magnitude whether using single or double precision long-range forces. The same cannot be said for the angular momentum, however, where the uncertainty for the FP32 simulations is higher than that for the FP64 simulations by more than an order of magnitude. Its time evolution is reminiscent of a Wiener process, { defined as a real-valued continuous-time stochastic process akin to one-dimensional Brownian motion \citep{Karatzas1991}.} Even so, the value of $\Delta L/L$ is still very low. As such, we conclude that statistically the outcomes between the two sets of 40 sims with different force precision calculations are the same. { Examples of the initial and final configuration of several simulations is depicted in Fig.~\ref{fig:tpfbe}, where we show the semi-major axis, eccentricity and masses of the planets and planetesimals.}

\subsection{Embryo simulations without gas}
{ The results of the simulations with planetary embryos and planetesimals without gas is given in Figs.~\ref{fig:embryosfinal} and \ref{fig:embryosevo}. The first figure displays the mass versus semi-major axis (top left), the spacing versus concentration (top-right), the relative changes in the energy (bottom-left), and the same for the angular momentum (bottom-right). The black dots are for simulations with FP64 long-range forces, and red are for FP32. In these simulations the values of $\Delta E/E$ are not continually rising, but show more of a random walk behaviour, with a similar magnitude for both single and double precision forces. This is different from that shown in Fig.~\ref{fig:chambers} where $\Delta E/E$ kept increasing.\\

Examples of the evolution of two simulations are shown in Fig.~\ref{fig:embryosevo}. The top four panels are for a run with FP64 forces, the bottom for a run with FP32 forces. The panels are: semi-major axis (top-left), eccentricity (top-right), mass (bottom-left), and inclination (bottom-right). The evolution of the two simulations appears similar when accounting for stochastic effects. The maximum masses, eccentricity, and inclinations are all of similar magnitude, as well as the time when most collisions finish (around 3 Myr).}

\begin{figure}
\resizebox{\hsize}{!}{\includegraphics{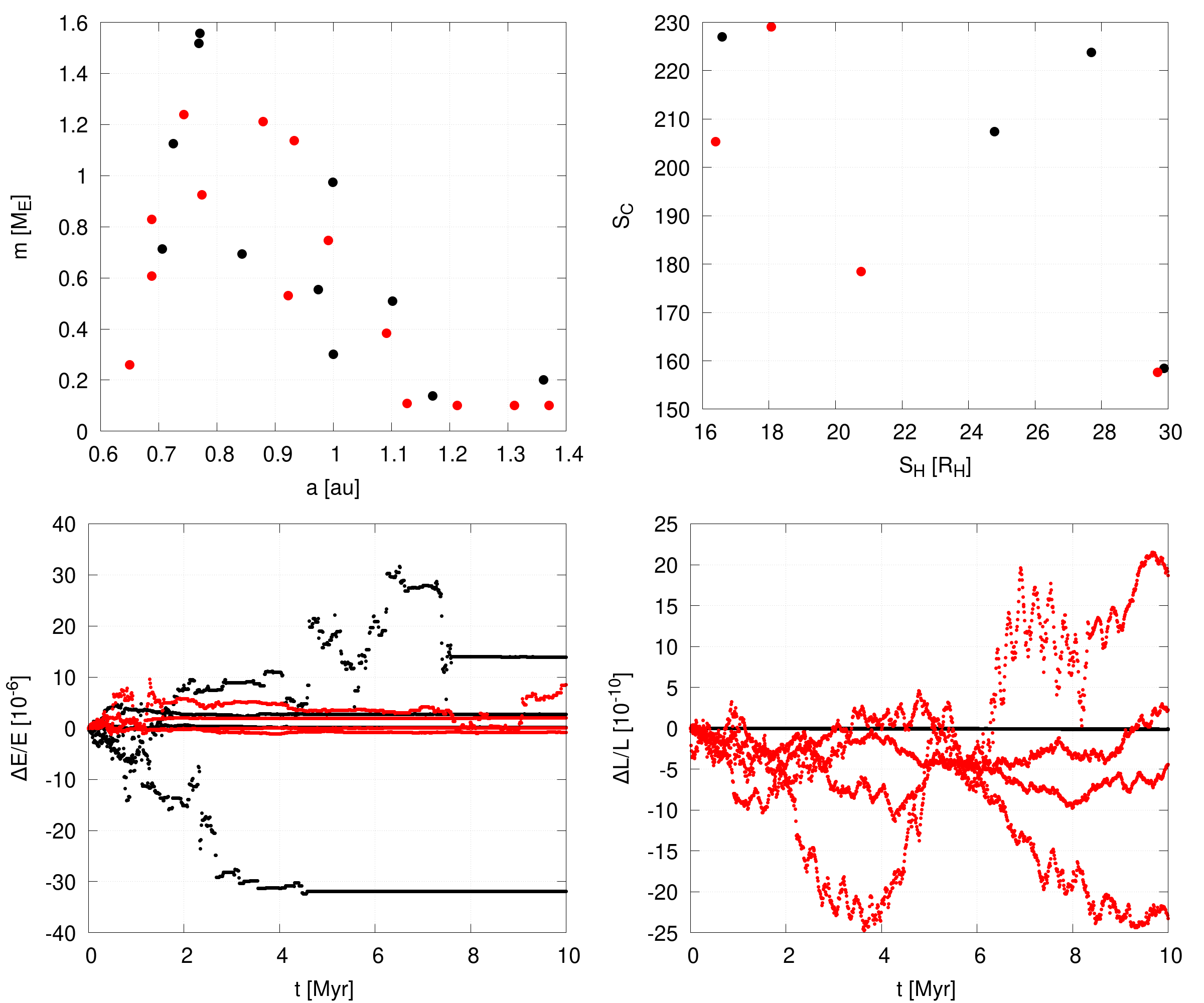}}
\caption{Output of the embryo experiment. The top-left panel depicts final mass versus semi-major axis, the top-right panel shows the concentration versus spacing parameters, the bottom-left panel is $\Delta E/E$, and the bottom-right panel is $\Delta L/L$. Black dots are for simulations run with FP64 forces, red for FP32.}
\label{fig:embryosfinal}%
\end{figure}

\begin{figure}
\resizebox{\hsize}{!}{\includegraphics{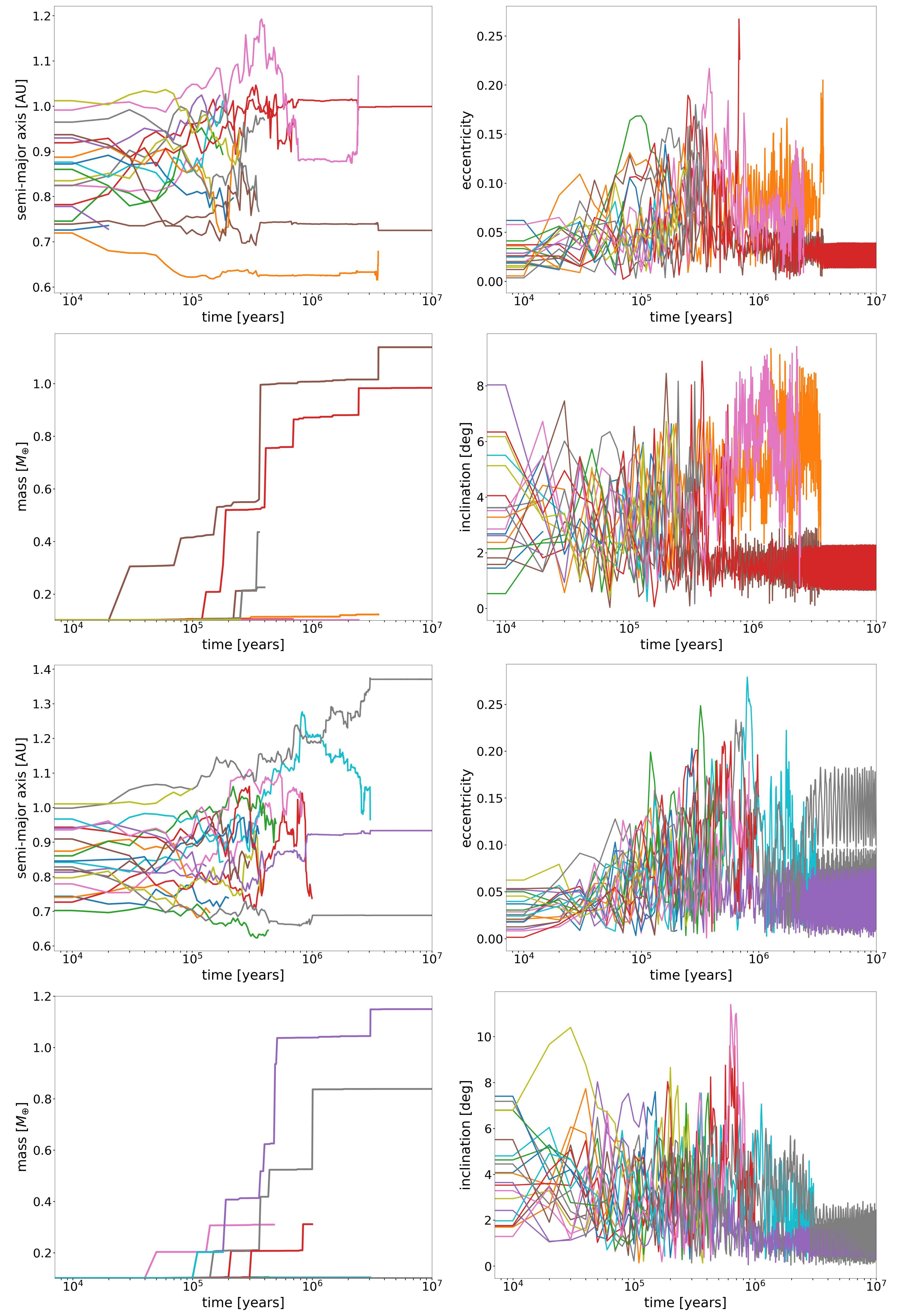}}
\caption{Embryo runs evolution examples. There are four panels per run. The top-left is the semi-major axis versus time, top-right is eccentricity versus time, bottom-left is mass versus time, and bottom-right is inclination versus time. The top four panels are an example of a simulation run with FP64 forces. The bottom four panels are for a simulation with FP32 forces.}
\label{fig:embryosevo}%
\end{figure}

\begin{figure}
\resizebox{\hsize}{!}{\includegraphics{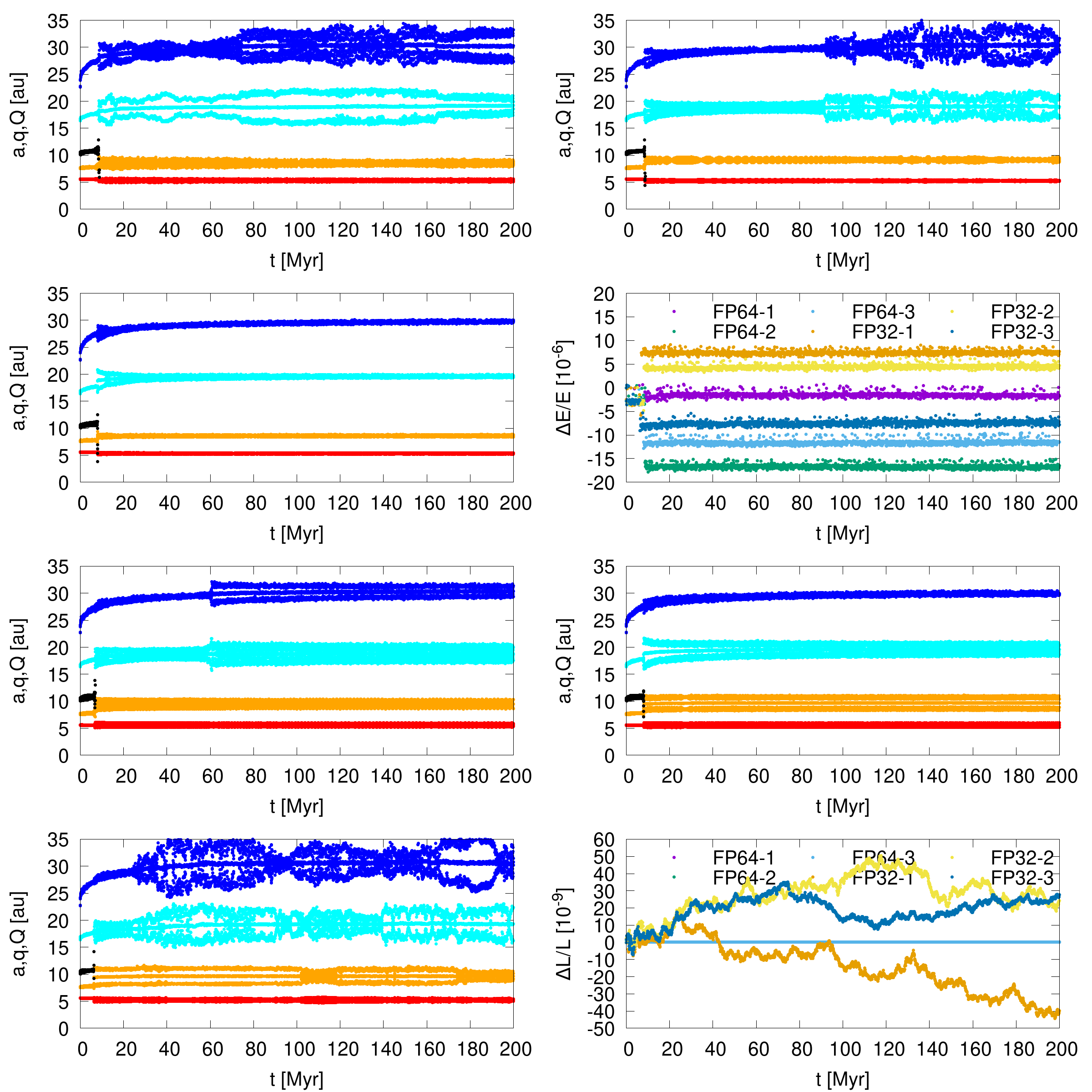}}
\caption{Evolution of giant planet dynamical instability runs, and corresponding energy and angular momentum uncertainties. The top three evolutionary tracks are for FP64 forces, while the bottom three are for FP32 forces. Red is for Jupiter, orange for Saturn, cyan for Uranus, and blue for Neptune. The evolution of the giant planets looks qualitatively similar irrespective of the force precision, and the energy uncertainties are also of similar magnitude. Once again the angular momentum conservation is better in FP64 mode. The high eccentricities in some runs of Uranus and Neptune have to do with the proximity to their 2:1 resonance.}
\label{fig:nice}%
\end{figure}

\subsection{Giant planet instability}
{The giant planet migration simulations were run with GENGA's TP2 mode, in which the giant planets feel the forces of the planetesimals, but the planetesimals do not feel each other \citep{grimmetal2022}. We have chosen to run the simulations in this mode to be able to compare them with output from the simulations of \citet{Wong2019,Wong2021} that were run with SyMBA \citep{Duncan1998}. \\

Due to the chaotic nature of these simulations the probability of the planets ending up near their current configuration is rather low \citep[e.g.][]{NM2012,BL2015}. Fortunately we found three simulations from each set where the planets' final configuration was reasonably close to the current system. We show the evolution of the giant planets' semi-major axis, perihelion, and aphelion in Fig.~\ref{fig:nice}, together with the relative uncertainties in the energy and angular momentum. The different evolution tracks shown for the giant planets are consistent with those reported in the literature \citep[e.g.][]{NM2012,Wong2019,Wong2021}. The values of $\Delta E/E$ are comparable for FP32 and FP64 again, and the jumps coincide with the ejection of the third ice giant around 5~Myr. The angular momentum in FP64 mode is once again extremely well conserved, while it shows a random walk in FP32 mode with $\vert \Delta L/L \vert <10^{-8}$. There is also no discernible difference in the migration speed of Neptune between the different simulations. One thing of note is that the FP32 simulations had on average 22\% fewer planetesimals remaining than the FP64 simulations, which could be due to the different evolution of the planets in each simulation.}

\subsection{Terrestrial planet formation with gas}
{ Figure~\ref{fig:chambersTPF} shows the mass-semimajor axis distribution, spacing versus concentration parameters, and relative uncertainties in energy and angular momentum for this experiment. Once again the black dots represent simulations with the long-range forces computed with FP64 precision, and red for FP32 precision. The influence of the gas lowers the overall angular momentum of the system because the forming planets migrate starwards. The larger changes in energy than in the previous simulations are caused by the dissipating gas disc. The large jump in one FP32 simulation is caused by a lunar-mass embryo being lost at the inner edge of the simulation. Generally the mass-semimajor axis distribution of both sets are similar, although the FP64 set has fewer bodies of Mars' mass and larger than the FP32 simulations.\\

We also notice that the systems in the FP64 simulations are generally sparser than the FP32 simulations. We do not have an adequate explanation for this difference. The planets in the FP32 simulations are, on average, 11\% more massive but are on average 0.7\% farther from the star, so this effect is likely not the result of migration. The average spacing for the FP64 simulations is $\langle S_H \rangle = 24.4 \pm 2.8$ and $\langle S_H \rangle = 20.8 \pm 1.5$ for FP32, indicating that they are consistent within 2$\sigma$. Similarly, the concentration parameters are $\langle S_C \rangle = 12.9 \pm 0.8$ for FP64 and $\langle S_C \rangle = 18.1 \pm 0.9$ for FP32, which are statistically inconsistent. In two of the FP64 simulations the most massive planet has $a>1$~au which does not happen in the FP32 simulations; this could explain the different concentration parameters between the two sets. Due to limited HPC access we chose not to run additional simulations to test whether this difference is an artefact of low-number statistics.\\

Figures~\ref{fig:evofp32} and~\ref{fig:evofp64} show snapshots of the evolution of one of the simulations. The output should be compared to what was reported by \citet{wooetal2021}: identically to their results we once again notice that the asteroid belt region is emptied out by the sweeping of the $\nu_5$ secular resonance as the gas disc dissipates. The evolution in the two figures looks qualitatively similar. The fact that the uncertainties in energy and angular momentum are quantitatively similar with both force prescriptions suggests to us that the different outcomes are due to stochastic effects within the simulations rather than due to the numerical methods employed.}

\begin{figure}
\resizebox{\hsize}{!}{\includegraphics{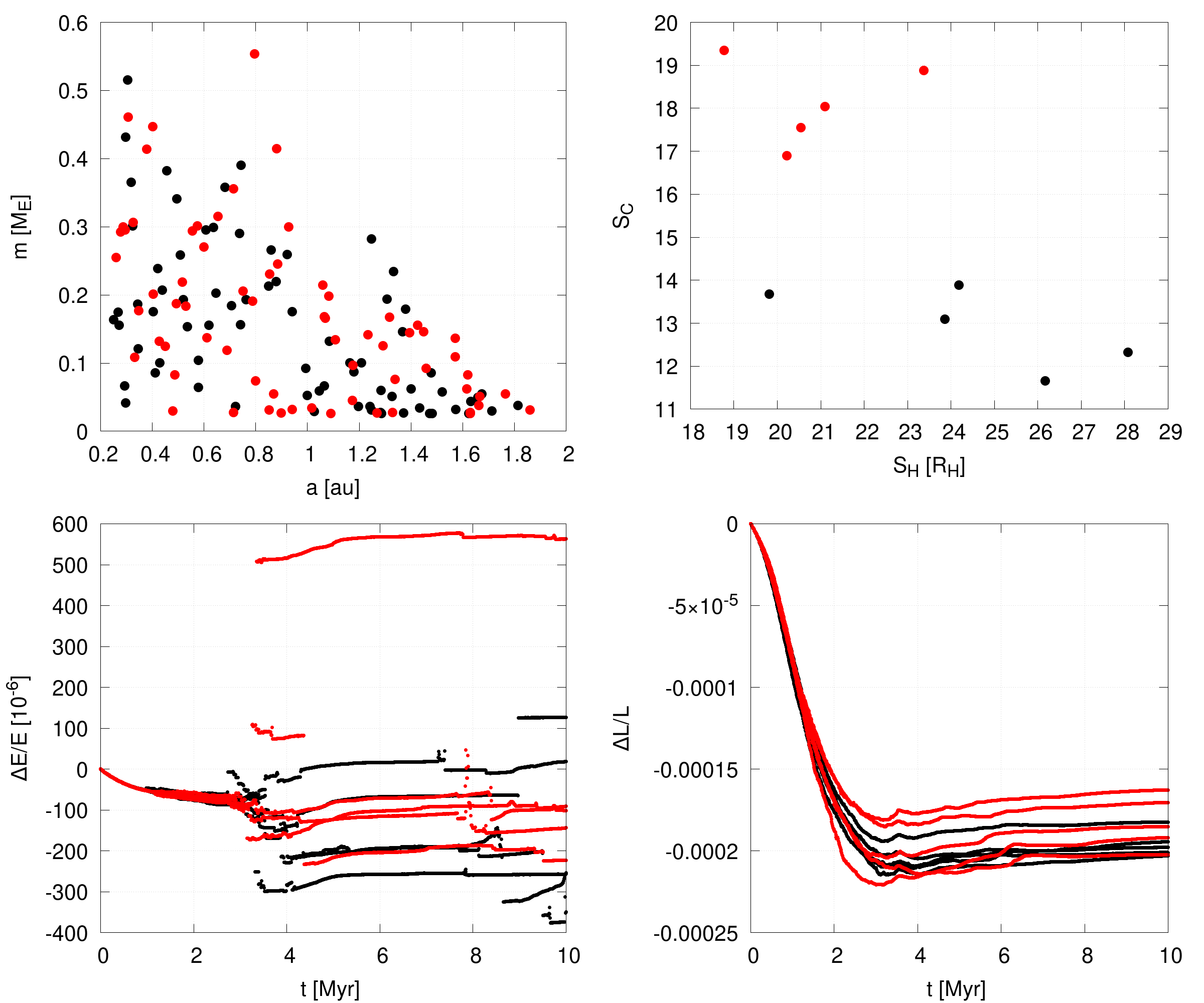}}
\caption{Output of the terrestrial planet experiment with gas. The top-left panel is final mass versus semi-major axis for all objects with a final mass $m>0.025$~$M_\oplus$, the top-right panel shows the concentration versus spacing parameters, the bottom-left panel is $\Delta E/E$, and the bottom-right panel is $\Delta L/L$. Black dots are for simulations run with FP64 forces, red for FP32.}
\label{fig:chambersTPF}%
\end{figure}

\begin{figure}
\resizebox{\hsize}{!}{\includegraphics{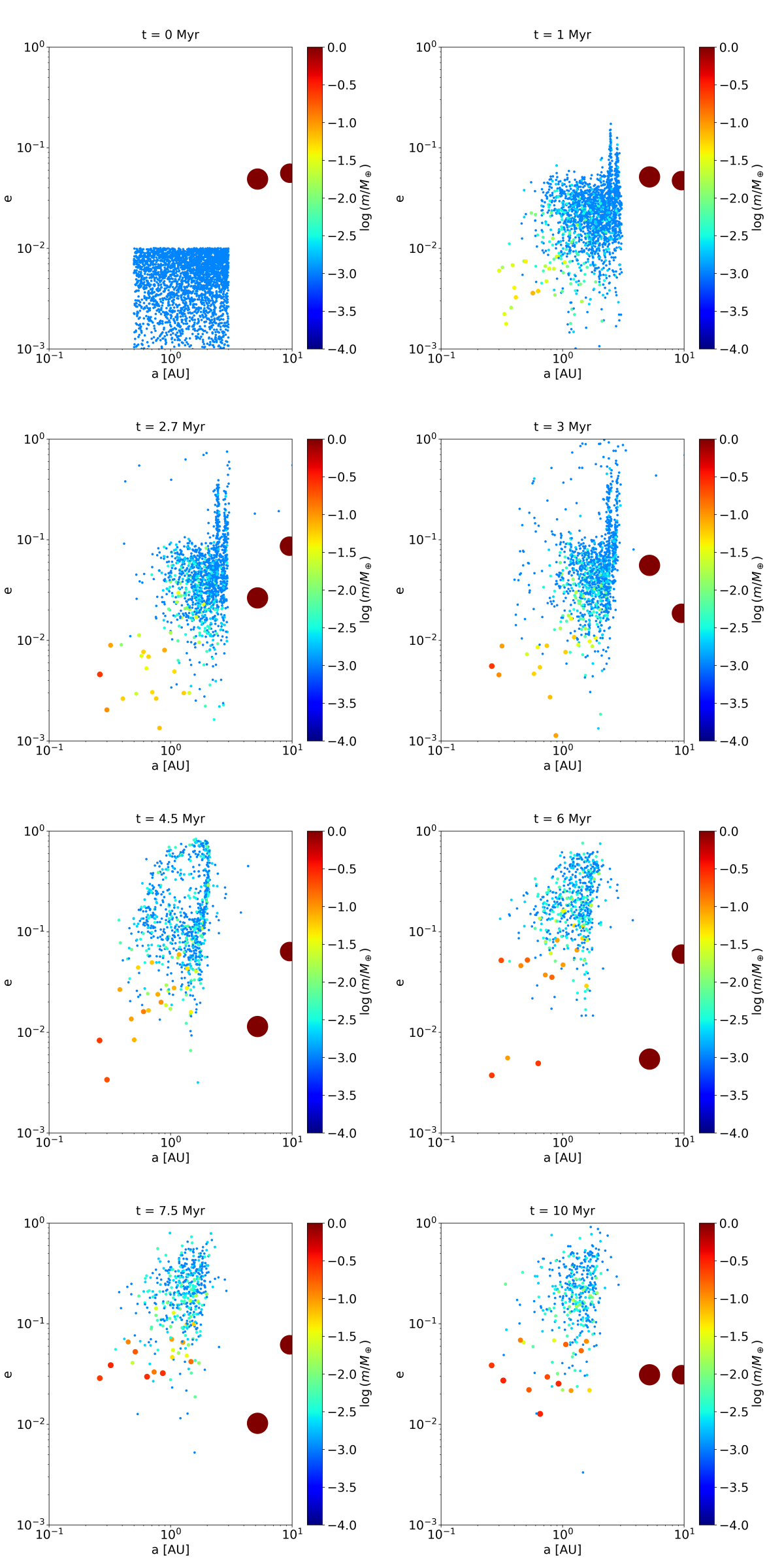}}
\caption{Snapshots of the evolution of a terrestrial planet formation simulation with gas disc. Panels show eccentricity versus semi-major axis with the colour coding being a proxy for the mass. The simulation was run with FP32 long-range forces.}
\label{fig:evofp32}%
\end{figure}

\begin{figure}
\resizebox{\hsize}{!}{\includegraphics{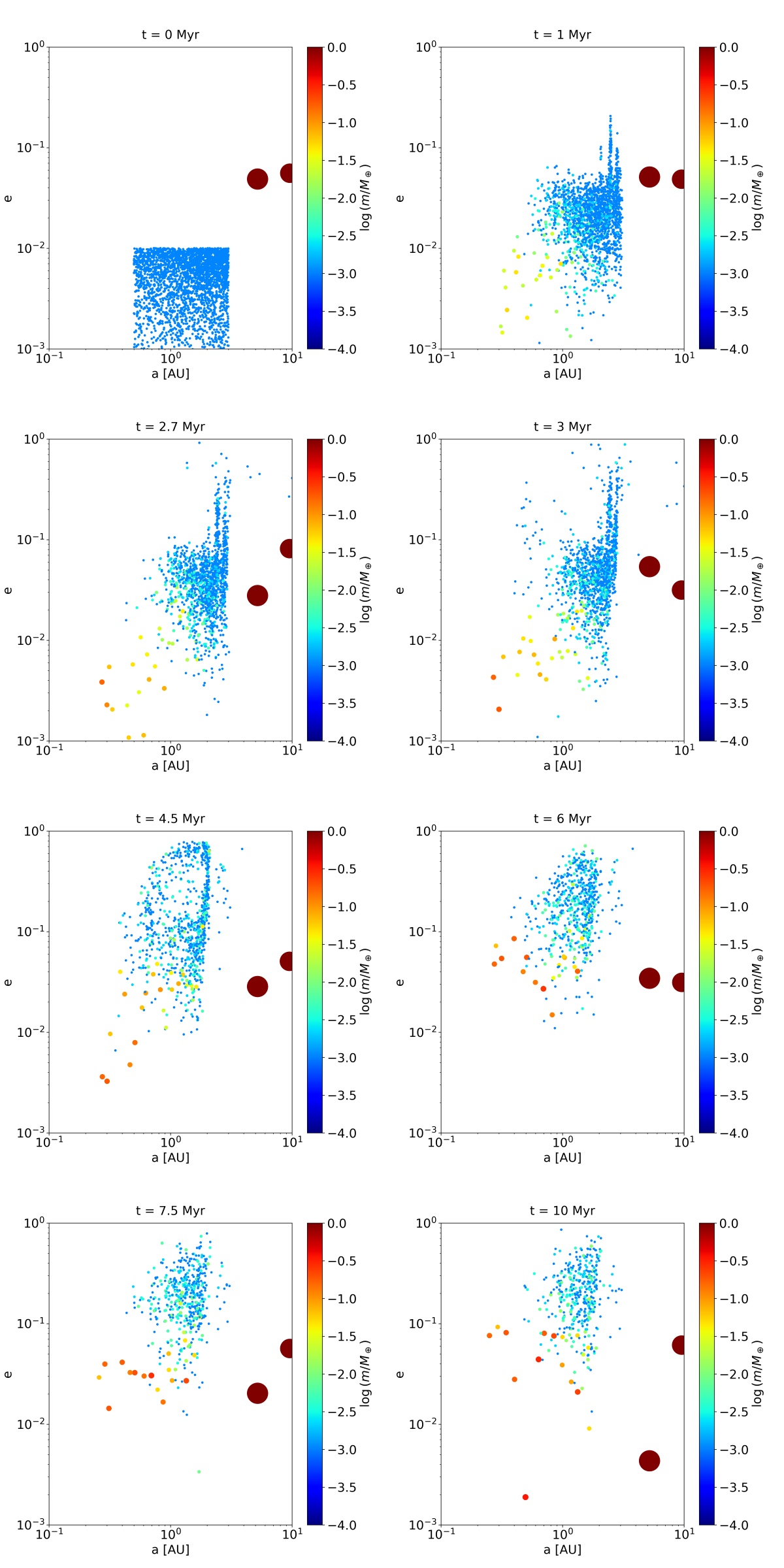}}
\caption{Snapshots of the evolution of a terrestrial planet formation simulation with gas disc. Panels show eccentricity versus semi-major axis with the colour coding being a proxy for the mass. The simulation was run with FP64 long-range forces.}
\label{fig:evofp64}%
\end{figure}

\subsection{High $N$ simulations on consumer-grade cards}
{ The previous experiments were meant to demonstrate that statistically the outcome of various different kinds of simulations are identical whether the long-range forces are computed in single or double precision. The only viable difference we have reported so far is that the relative change in the angular momentum is less well preserved when using FP32 forces. \\}

We now turn to the high-N simulation without gas that was run on different GPUs, including several consumer-grade cards. { Figure~\ref{fig:highres} shows a snapshot of the beginning and end of two of these simulations run on the 1080 Ti. The left column was run with FP32 precision and the right column with FP64 precision. There is no obvious difference between the results.} In Fig.~\ref{fig:sps} we show the number of steps per second as a function of the total number of steps for some of the cards that we tested. We show the results for tuned (top row) and untuned (bottom row) simulations in FP32 (left column), and FP64 (right column) precision. The speedup for the tuning is mostly apparent in the first $\sim$2 million steps. For the non-tuned simulations the number of steps per second begins as low as 0.32 on most cards, and rises rapidly, { while for the tuned simulations this never drops below $\sim$2 steps per second}. The reason for this initially slow performance is that there are many planetesimals that are forming almost one entire large close encounter group, so GENGA spends all of its time in the Bulirsch-Stoer kernels \citep{grimmetal2022}, which are slow by nature, and are also always performed in FP64 precision. Once most encounters are resolved through collisions after about 2 million steps the code spends most of its time doing the force calculations, { and the simulations speed up by more than an order of magnitude.} \\

The difference in speed between the FP32 and FP64 calculations on the consumer-grade cards is evident, { and is typically a factor of four when not performing close encounter calculations}. For the tuned simulations the Bulirsch-Stoer kernel is called much less frequently and the number of steps per second does not drop much below two for FP64, and eight for FP32 on the consumer-grade cards. The overall speedup across the whole simulation from the tuning is roughly a factor of two, purely due to the speeding up of the close encounter phase in the first 2 million steps.\\

\begin{figure}
\resizebox{\hsize}{!}{\includegraphics{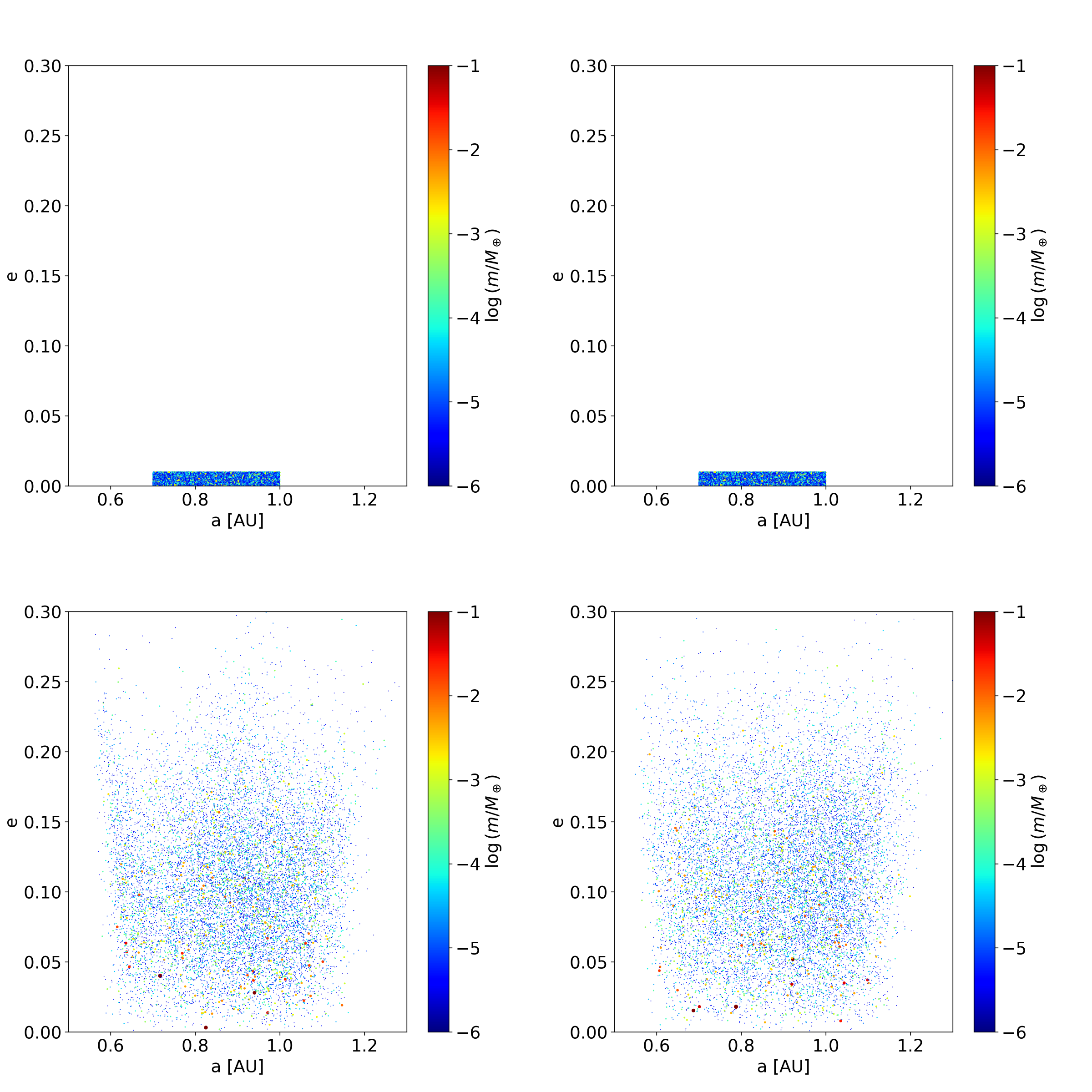}}
\caption{Snapshots of semi-major axis, eccentricity, and mass at the beginning (top row) and end (bottom row) of tuned high resolution simulations run on the 1080 Ti. The left column was run with single precision long-range forces and the right column in double precision.}
\label{fig:highres}%
\end{figure}

\begin{figure}
\resizebox{\hsize}{!}{\includegraphics{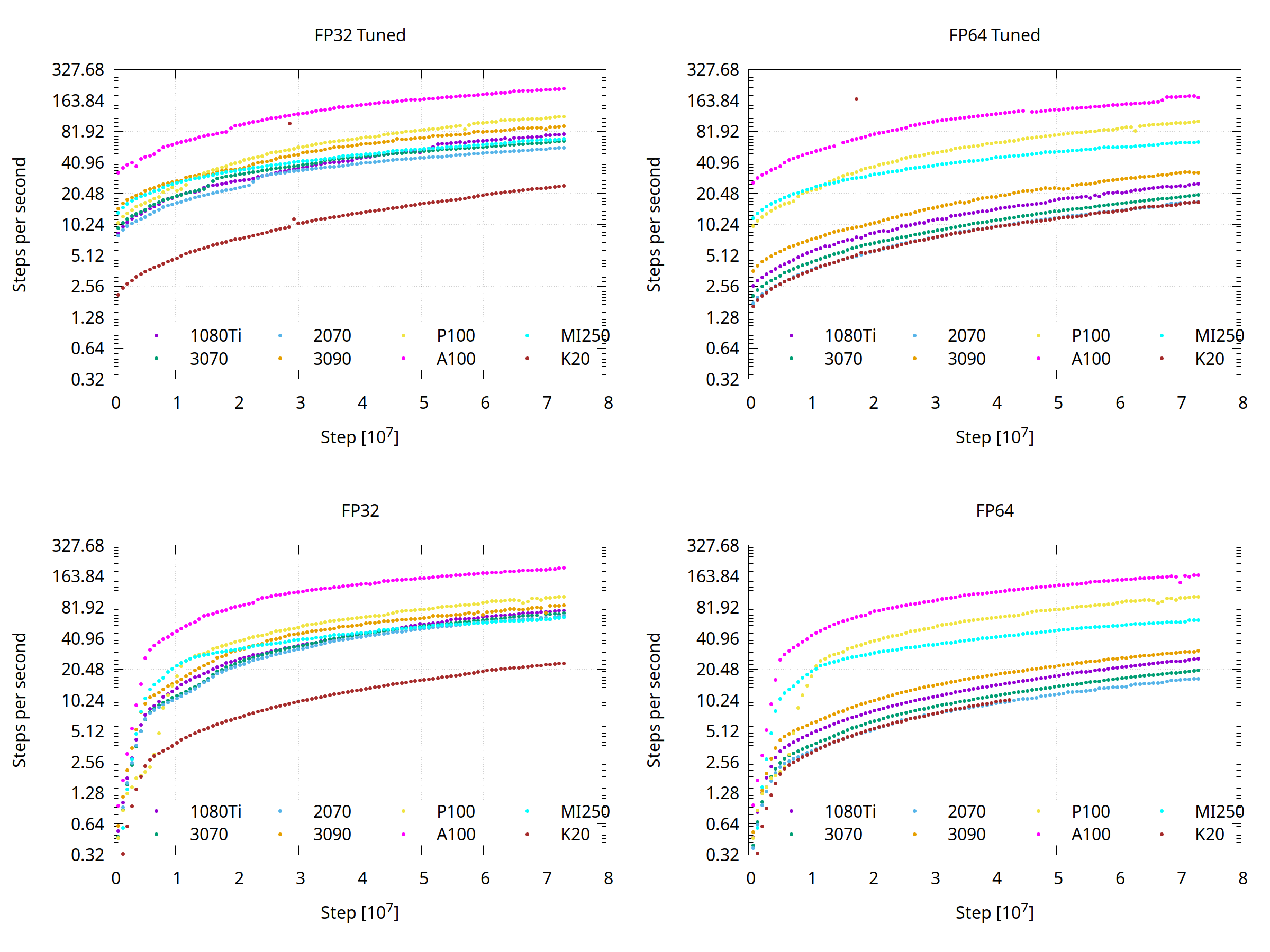}}
\caption{Number of steps per second for various GPUs. The top two panels are for the tuned simulations, while the bottom two panels are for untuned simulations. There is a large difference between sims with FP64 and FP32 precision forces on consumer-grade cards (3000, 2000 and 1000 series). The tuning speeds up the first 10-20 million steps when there are many close encounters.}
\label{fig:sps}%
\end{figure}

\begin{figure}
\resizebox{\hsize}{!}{\includegraphics{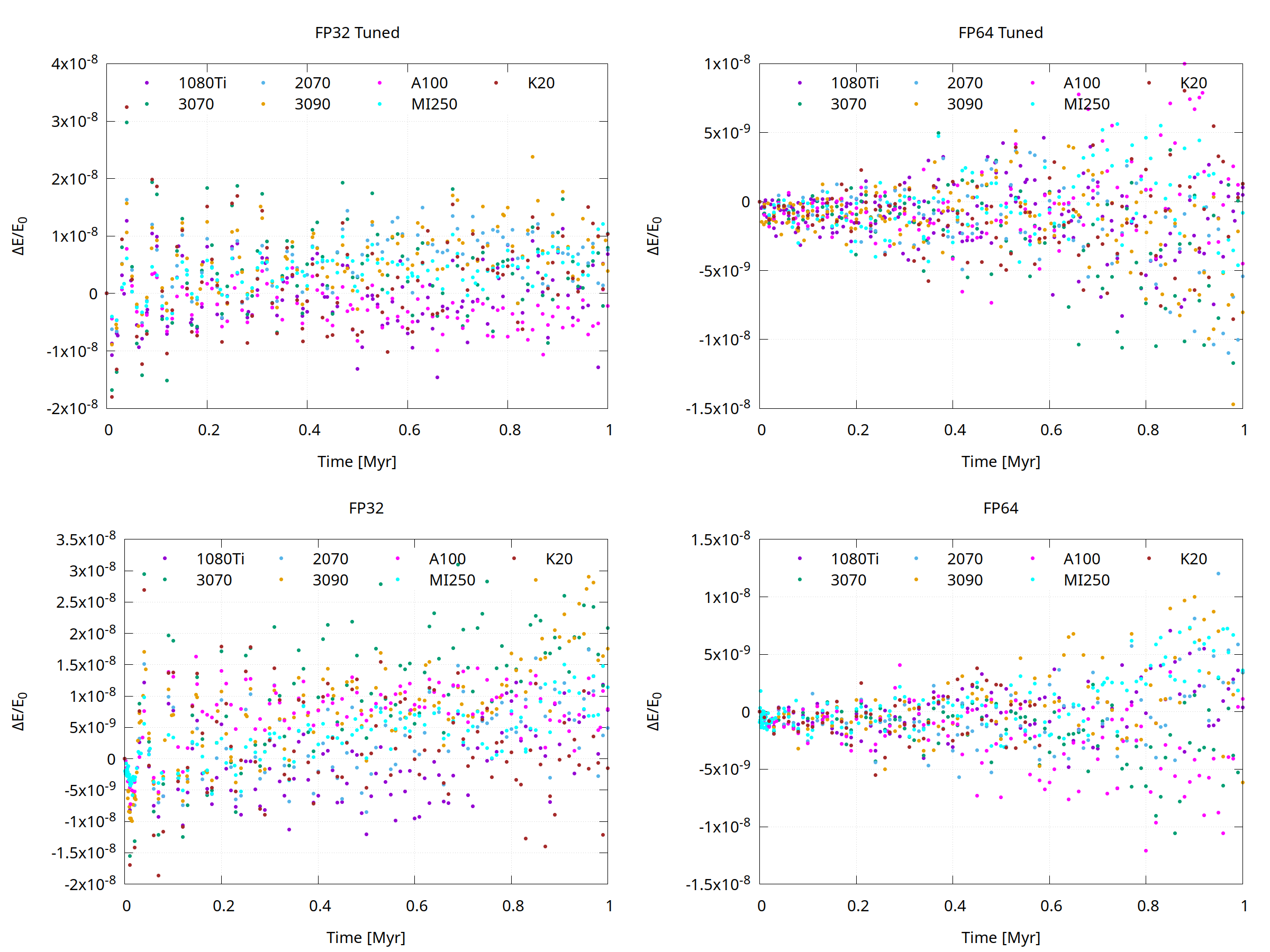}}
\caption{Fractional change in total energy on various GPUs. The top row depicts the tuned simulations. The fluctuations are comparable for simulations run with FP32 precision forces than with FP64 precision.}
\label{fig:de}%
\end{figure}

\begin{figure}
\resizebox{\hsize}{!}{\includegraphics{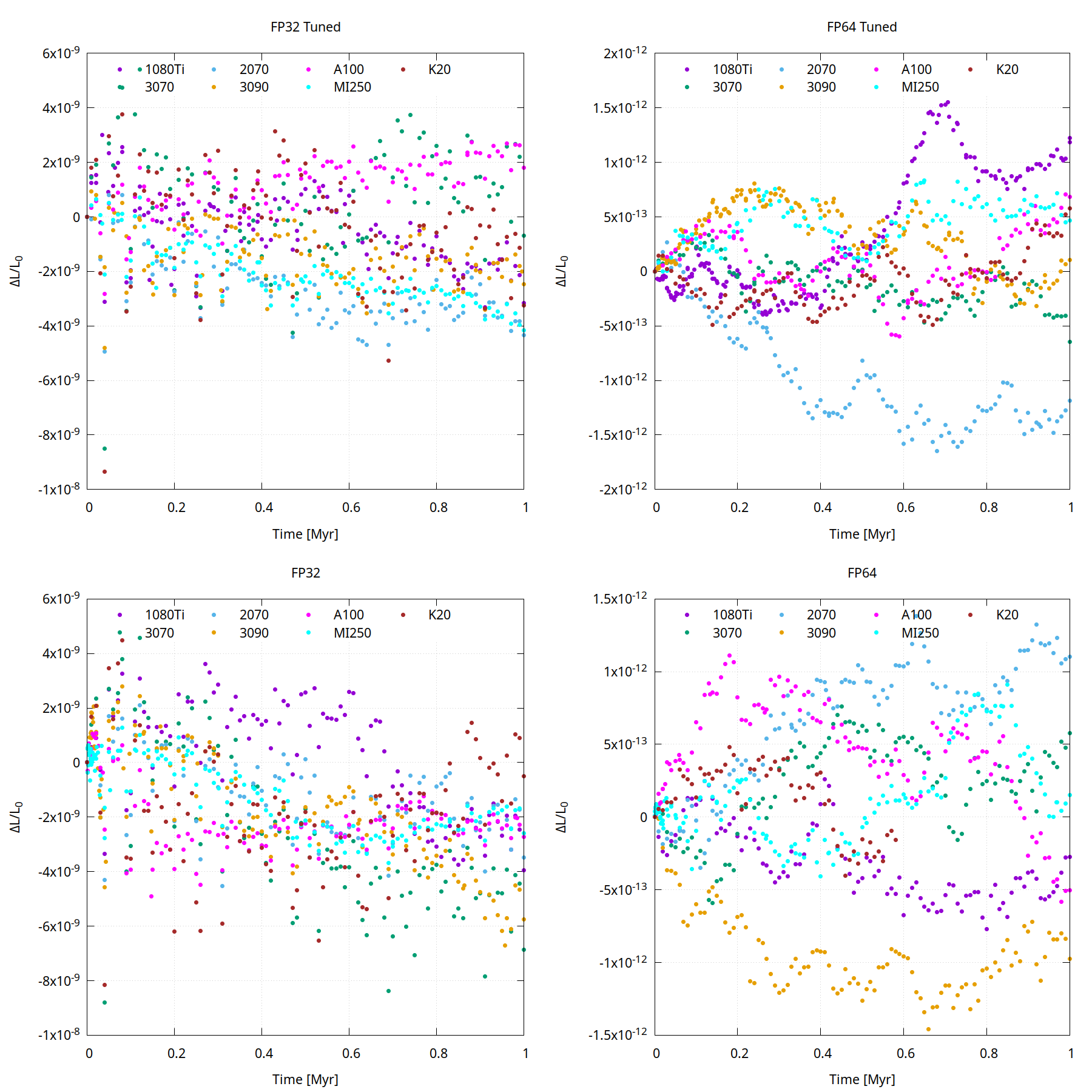}}
\caption{Fractional change in total angular momentum on various GPUs. The top row depicts the tuned simulations. The simulations with FP64 forces seem to preserve the angular momentum at least two orders of magnitude better than the simulations with FP32 forces.}
\label{fig:dl}%
\end{figure}

Note that after 2 million steps the FP32 simulations on all consumer-grade cards are approximately 25\% to 50\% of the speed of the A100, which are comparable to the older P100, while the FP64 simulations are a factor of 3-4 slower, { in other words only about 10\%. After the close encounter phase the tuning does not seem to generate any significant speed increases.}\\

At the end there are roughly 12\,000 planetesimals left for all simulations. At that time the 1080 Ti does 908\,000 steps per second per particle in FP32 precision and 309\,000 in FP64 precision, a difference of a factor of three. The other consumer-grade cards have similar ratios. For the A100 GPU the number of steps per second per particle is about 2.2-2.5 million in both FP64 and FP32 respectively; the ratio is very close to one. The difference between the 1080 Ti and the A100 corresponds well with their theoretical peak FP32 performance ratio, which suggests that by calculating the forces in FP32 precision on the 1080 Ti we tap into its computing capabilities to the same degree as we do on the Tesla A100. The AMD MI250, on the other hand, compares unfavourably to the Tesla P100 and A100, { and is comparable in speed to the} consumer-grade cards when they execute FP32 force precision. The reason for this is that the HIP version of GENGA is not as well developed as the intrinsic CUDA version, so there is the potential to harvest more speed from the MI250 cards.\\

Figs.~\ref{fig:de} and~\ref{fig:dl} show the relative deviations in energy and angular momentum the same different types of GPUs that we tested. It is clear that the magnitude of the deviations in both quantities is similar for all cards, but just as in the previous experiments there is a difference whether the forces are calculated in FP32 or FP64 precision. { Just as with the previous experiments} the penalty in the accumulated errors in energy in the FP32 simulations are comparable to those simulations running in FP64 precision, while the deviations in angular momentum are much larger in FP32 precision than in FP64 precision.\\

Nevertheless, we advocate running GENGA in FP32 precision on consumer-grade cards. In Table~\ref{tab:complete} we show the time it takes to complete the simulations on the various cards with various precision and tuning settings. { The untuned:tuned ratio of simulation duration in FP64 mode ranges from 1.4 for the 1080 Ti to 3.7 on the P100, while the range is 2.3 on the 2070 to 4.1 on the P100 in FP32 mode. The time FP64T:FP32T for the consumer grade cards ranges from 3.4 on the 3090 to 4.2 on the 2070. As such, running tuned simulations with FP32 precision forces on consumer-grade cards increases the calculation speed by approximately an order of magnitude over untuned simulations with FP64 precision long-range forces. Even though running in FP32 precision increases power usage, overall the total energy used to run the entire simulation is less than using FP64 precision.}

\begin{table}
\begin{tabular}{r|cccc}
GPU & FP32 (d) & FP32T (d) & FP64 (d) & FP64T (d) \\ \hline \\
1080 Ti & 60.6 & 26.6 & 121.5 & 87.4 \\
2070 & 69.4 & 30.6 & 173.3 & 129.5\\
3070 & 68.2 & 25.9 & 201.4 & 109.9\\
3090 & 51.4 & 19.0 & 101.7 & 65.5 \\
K20 & 193.2 & 97.7 & 220& 131.1\\
MI250 & 78.7 & 22.0 & 80.7 & 24.1\\
P100 & 77.6 & 19.1 & 77.7 & 21.0 \\
A100 & 27.5 &8.5 & 28.8 & 9.3
\end{tabular}
\caption{Completion time in days for the various cards we tested. The T suffix means the tuned simulations. The drastic decrease in running time between the tuned and non-tuned simulations is caused by the tuning speeding up the first 10\% of the simulation.}
\label{tab:complete}
\end{table}

\section{Conclusions}
We showed that computing the mutual long-range forces between planetesimals in FP32 precision in GENGA speeds up the computation on consumer-grade cards by a factor of 3-4. It does not lead to larger accumulations in errors in relative energy { for several different numerical experiments}, but it does cause { much higher} errors in angular momentum conservation. { Nevertheless, we argue that the angular momentum errors are all within acceptable limits. This behaviour holds across various different experiments.} \\

{ By performing the long-range forces in FP32 precision, GENGA's performance on consumer-grade cards is approximately a factor of three faster than when the whole simulation is performed only in FP64 precision, and a factor of three slower than the Tesla A100. It also reduces overall energy usage per simulation.} \\

The option of using single precision gravitational forces can be enabled by setting \texttt{Do Kick in single precision = 1} in the parameter file of GENGA.

\section*{Acknowledgments}
We acknowledge EuroHPC Joint Undertaking for awarding us access to Vega at IZUM, Slovenia. We further acknowledge resource usage on LUMI at CSC, Finland, and Betzy operated by Sigma2 in Norway as a part of Stephanie C. Werner’s project NN9010K. Some calculations were performed on UBELIX (\url{http://www.id.unibe.ch/hpc}), the HPC cluster at the University of Bern. This study is supported by the Research Council of Norway through its Centres of Excellence funding scheme, project No. 223272 CEED (PH) and 332523 PHAB (PH). GENGA can be obtained from \url{https://bitbucket.org/sigrimm/genga/}

\bibliographystyle{aa}
\bibliography{References}
\end{document}